\documentclass[aps,prl,twocolumn,showpacs,superscriptaddress,groupedaddress]{revtex4-1}

\newcommand{\ket}[1]{\left|#1\right\rangle}

\newcommand{\sandwich}[3]{\left \langle #1 | #2|#3\right\rangle}
\newcommand{\Ca}{Ca$^{+}$}
\newcommand{\Ba}{Ba$^{+}$}
\newcommand{\Ra}{Ra$^{+}$}
\newcommand{\Yb}{Yb$^{+}$}
\newcommand{\Sr}{Sr$^{+}$}
\newcommand{\Hg}{Hg$^{+}$}
\newcommand{\Sc}{Sc$^{++}$}
\newcommand{\avg}[1]{\langle #1 \rangle}

\newcommand{\branching}{$0.93565(7)$}
\newcommand{\branchingsm}{$0.06435(7)$}

\usepackage[utf8x]{inputenc}
\usepackage{graphicx}
\usepackage{dcolumn}
\usepackage{bm}
\usepackage{amsmath}
\usepackage{amssymb}
\usepackage{subfigure}
\usepackage{subfloat}

\begin{document}

\title{Precision Measurement Method for Branching Fractions of Excited P$_{1/2}$ States Applied to $^{40}$\Ca}

\author{Michael Ramm, Thaned Pruttivarasin, Mark Kokish, Ishan Talukdar, Hartmut H\"affner}

\affiliation{Department of Physics, University of California,  Berkeley, CA 94720, USA}

\email{hhaeffner@berkeley.edu}
\date{\today}

\begin{abstract}

We present a method for measuring branching fractions for the decay of $J = 1/2$ atomic energy levels to lower-lying states based on time-resolved recording of the atom's fluorescence during a series of population transfers. We apply this method to measure the branching fractions for the decay of the 4$^{2}$P$_{1/2}$ state of $^{40}$\Ca to the 4$^{2}$S$_{1/2}$ and 3$^{2}$D$_{3/2}$ states to be \branching\ and \branchingsm, respectively. The measurement scheme requires that at least one of the lower-lying states be long-lived. The method is insensitive to fluctuations in laser light intensity and magnetic field and is readily applicable to various atomic species due to its simplicity. Our result distinguishes well among existing state-of-the-art theoretical models of \Ca.

\end{abstract}

\maketitle

\section{Introduction}

Comparing theoretically calculated and experimentally measured observables of complex atoms provides a crucial step towards a complete understanding of atomic structure. Atomic properties such as excited state lifetimes and branching fractions of energy levels in various atomic species are of special interest because they can be readily measured both in atomic ensembles \cite{Hannaford, Fivet} and in single atoms \cite{Kurz, Gerritsma}. Knowledge of these atomic properties is especially important for	studying spectra of radiation from stars and other galactic objects \cite{Barbuy, Curtis}. Furthermore, precise measurement of atomic lifetimes and branching fractions can also be used to probe effects of weak interactions such as parity non-conservation of atomic levels \cite{Fortson,Pal,Versolato,Nguyen}.

Precision measurements with trapped ions have benefited from many technological developments in the field of quantum information processing \cite{Roos, Rosenband}. Specifically, the lifetimes of the long-lived $D$ states \cite{Kreuter} and the branching fractions for the decay of the 4$^{2}$P$_{3/2}$ state in calcium ions (\Ca) \cite{Gerritsma} and barium ions (\Ba) \cite{Kurz} have been measured with high precision. Both measurements of the branching fractions use narrow line-width lasers to address the quadrupole transition and perform state readout, making it challenging to apply the method to other atomic species. 

In this work, we present a simple method to measure the branching fractions for the decay of the excited states with $J=1/2$. Due to the simplicity of the experimental setup, the method is readily applicable to many atomic species. We show that many of the systematic uncertainties such as laser light polarization and magnetic field fluctuations can be disregarded. We apply this method to perform a precision measurement of the branching fractions for the decay of the 4$^{2}$P$_{1/2}$ state to the 4$^{2}$S$_{1/2}$ and the 3$^{2}$D$_{3/2}$ of $^{40}$\Ca.

\section{Theory}

\begin{figure}
\begin{center}
\includegraphics[width = 0.4\textwidth]{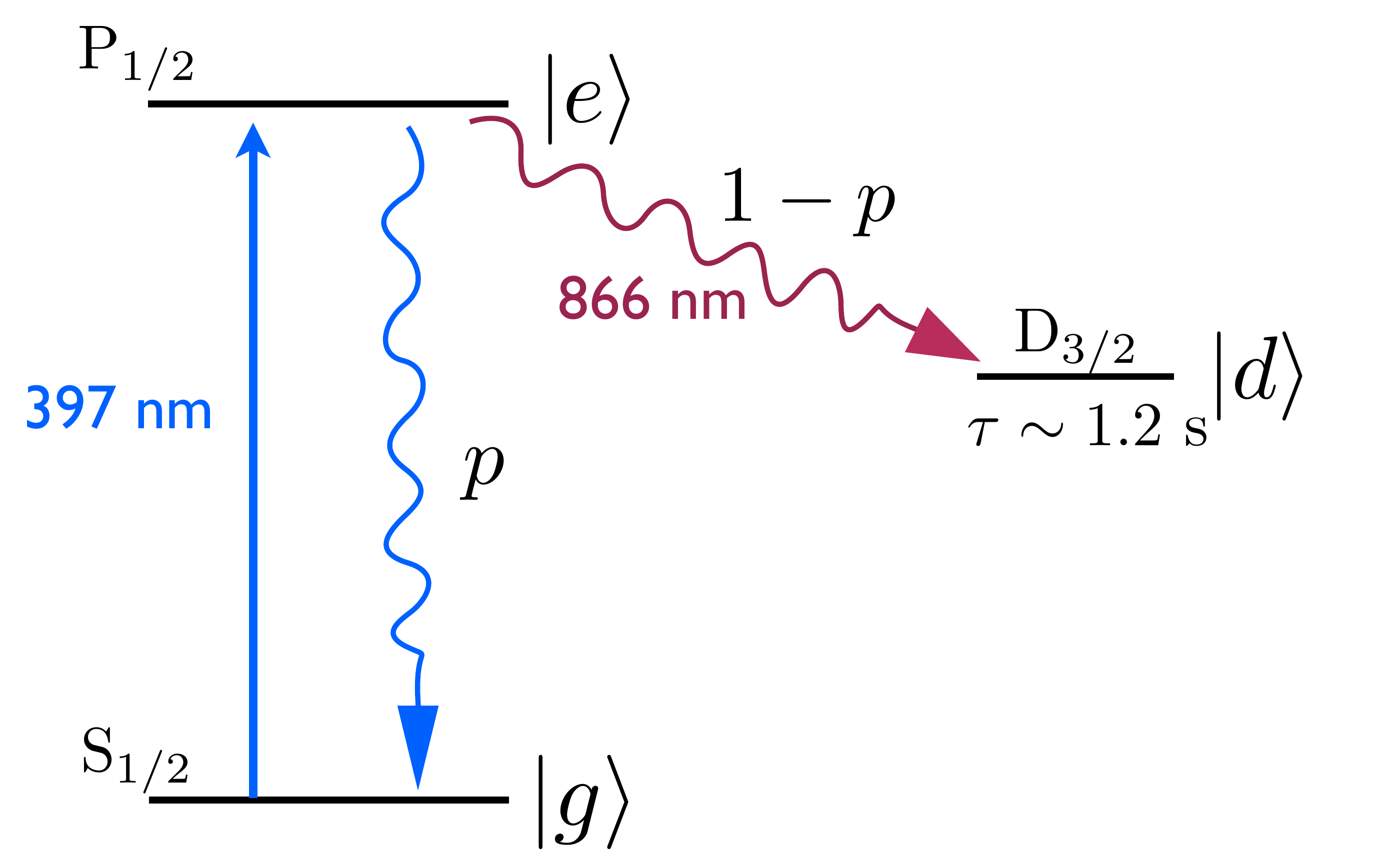}
\caption{\label{diagram} Energy diagram of $^{40}$\Ca. The states considered in our analysis are the 4$^{2}$S$_{1/2}$, 4$^{2}$P$_{1/2}$ and 3$^{2}$D$_{3/2}$, which we label as $\ket{g}$, $\ket{e}$ and $\ket{d}$, respectively. The ion in $\ket{e}$ can either decay back to $\ket{g}$ with probability $p$ or to $\ket{d}$ with probability $1-p$. Diode lasers at 397 nm and 866 nm are used for Doppler cooling.}
\end{center}
\end{figure}

In this section, we show that the branching fractions for the decay of the excited state can be measured by simply counting the number of photons scattered from the atom. We consider an atomic energy level scheme where the excited state, $\ket{e}$, can decay either to a ground state, $\ket{g}$, or to a long-lived state, $\ket{d}$, with probabilities of $p$ and $1-p$, respectively (as shown in Figure \ref{diagram} for $^{40}$\Ca). With laser light addressing the  $\ket{g} \rightarrow \ket{e}$ transition present, (397 nm for $^{40}$\Ca), the atom goes through many scattering events before it eventually ends up in $\ket{d}$ and the scattering stops. We prove that the mean number of spontaneously emitted photons per atom during this population transfer to $\ket{d}$ only depends on $p$ and is given by $\avg{n} =1/(1-p)$.

We use optical Bloch equations \cite{Cohen} to treat the interaction of the atom with laser light. We model the atom as a two-level system consisting of $\ket{e}$ and $\ket{g}$ with an irreversible loss of the excited state population to a long-lived state $\ket{d}$. Starting from $\dot{\rho}(t) = [H(t),\rho(t)]$, where $\rho(t)$ is the density matrix and $H(t)$ is the Hamiltonian of the system, for each component of the density matrix, $\rho_{ij}(t)=\sandwich{i}{\rho(t)}{j}$, we have
\begin{subequations}
\begin{eqnarray}
\dot{\rho}_{ee}(t) &=& \frac{i\Omega(t)}{2}(\rho_{eg}(t)-\rho_{ge}(t))-\Gamma\rho_{ee}(t)\\
\dot{\rho}_{gg}(t) &=& \frac{i\Omega(t)}{2}(\rho_{ge}(t)-\rho_{eg}(t))+p\Gamma\rho_{ee}(t)\\
\dot{\rho}_{eg}(t) &=& (i\Delta(t)-\frac{\Gamma}{2})\rho_{eg}(t)+\frac{i\Omega(t)}{2}(\rho_{ee}(t)-\rho_{gg}(t))\ \ \ \ \\
\dot{\rho}_{dd}(t) &=& (1-p)\Gamma\rho_{ee}(t), \label{eq:rate}
\end{eqnarray}
\end{subequations}
where $\Omega(t)$ is the Rabi frequency associated with the intensity of the laser, $\Delta(t)$ is the detuning of the laser and $\Gamma$ is the decay rate of the excited state. 

The mean number of spontaneously emitted photons, $\avg{n}$, can be calculated from
\begin{eqnarray}
\avg{n}&=&\int_0^{\infty}\Gamma\rho_{ee}(t)dt \\
&=& \frac{1}{1-p}\int_0^{\infty}\dot{\rho}_{dd}(t) dt =\frac{1}{1-p},
\end{eqnarray}
where we substitute in the expression for $\rho_{ee}(t)$ from Eq (\ref{eq:rate}) and use the boundary value of $\rho_{dd}(t=0) = 0$ and $\rho_{dd}(t\rightarrow\infty) = 1$. The last photon is emitted from the $\ket{e}$ to $\ket{d}$ transition. Hence, the number of photons spontaneously emitted from $\ket{e}$ to $\ket{g}$ is given by

\begin{equation}
\avg{N} = \avg{n} - 1=\frac{p}{1-p}.
\end{equation}

Since $\avg{N}$ does not contain any explicit form of $\Omega(t)$ or $\Delta(t)$, it is independent of light intensity and frequency fluctuations. Additionally, any Doppler shifts due to the temperature of the atom or its motion (such as micro-motion in a Paul trap or cyclotron motion in a Penning trap) do not affect $\avg{N}$.

If the excited state has $J=1/2$, the photon emission pattern is isotropic as long as the photons are detected regardless of their circular polarizations \cite{Budker}. Particularly, the Hanle effect will have no influence on photon counting \cite{Gallagher1963}. As a result, $\avg{N}$ is not sensitive to laser light polarization and the magnetic field. In practice, a finite magnetic field and linear polarization of the laser light should be used to avoid any optical pumping and coherent population trapping. Such effects can reduce $\Omega(t)$ and lead to significant systematic uncertainties due to the finite measurement time of the population transfer.

\section{Experimental setup and procedure}

We measure the branching fractions of the 4$^{2}$P$_{1/2}$ level of $^{40}$\Ca by performing a series of population transfers. As shown in the previous section,  the branching fraction, $p$ can be obtained by counting the the number of photons emitted at 397 nm as the ion is pumped from the 4$^{2}$S$_{1/2}$ to the  3$^{2}$D$_{3/2}$ state. The transfer is repeated many times in order to precisely measure the value of $\avg{N}$ and, hence, calculate $p$.

We trap a crystal of 13 ions in a linear Paul trap with an inter-electrode distance of 1.0 mm. (While a larger crystal reduces the measurement time, we choose the number of ions such that the ions remain crystallized throughout the measurement process.) We apply a radio frequency (RF) voltage of 200 $V_{\text{pp}}$ with a frequency of 30 MHz to the nearest electrodes. One pair of the electrodes is driven 180 degrees out of phase with respect to the other pair. Together with 2 V DC applied to each of the end-cap electrodes, we obtain confinement frequencies of 1 MHz and 100 kHz in the radial and axial directions, respectively. 

The ions are cooled close to the Doppler limit with a red-detuned laser at 397 nm and a repumping laser at 866 nm. Both lasers are linearly polarized and pointed at 45 degrees with respect to all the trap axes in order to provide sufficient cooling of all motional modes. In our typical measurement, we apply a magnetic field of 1.0 G along the trap axis to lift the degeneracy of the Zeeman sublevels. 

The fluorescence from the ions is detected along the axis perpendicular to both the laser and the magnetic field. We use a photomultiplier tube (PMT) with a pair of interference filters (Semrock Brightline\textsuperscript{\textregistered} FF01-414/46-25 and FF01-377/50-25) to detect photons only at 397 nm. The filters suppress the fluorescence at 866 nm by at least 100 dB. The detection efficiency for the setup is $1.2 \times 10^{-3}$.

The external-cavity diode lasers at 397 nm and 866 nm are locked to temperature controlled Fabry-Perot cavities in order to stabilize their frequencies. We pass the 397 nm laser light through two dispersive prisms (Thorlabs PS850) to filter out background spontaneous emission which may excite the ion to the 4$^2$P$_{3/2}$ state. This background absorption is reduced to less than one event per 20 minutes for each ion. Both lasers are passed through acousto-optic modulators (AOMs) in a double-pass configuration. During the pulse sequences, the laser light is switched off using a direct digital synthesizer (DDS) and an RF switch to interrupt the voltage supplied to the AOM. This achieves a light extinction ratio of better than 70 dB. 

The procedure for measuring the branching fractions is as follows: we start with all the ions in the ground state and turn on the laser light at 397 nm (pump pulse) to transfer all ions to $\ket{d}$. The number of detected photons during the pump pulse is $\epsilon \avg{N}$ per ion during where $\epsilon$ is the detection efficiency. Once the ions are in $\ket{d}$, applying laser light at 866 nm (reset pulse), brings the ions back to the ground state. During this process, exactly one blue photon per ion is emitted, so on average, $\epsilon$ photons per ion are detected. By repeating these transfers $T$ times, we detect $N_b = \epsilon T\avg{N}=\epsilon Tp/(1-p)$ blue photons from the pump pulse and detect $N_r = T\epsilon$ blue photons during the reset pulse. Then the branching fraction $p$ is given by the ratio:
\begin{equation}
\label{p_equation}
p=\frac{N_b}{N_r+N_b},
\end{equation}
which is independent of the detection efficiency $\epsilon$.

\begin{figure}
\begin{center}
\includegraphics[width = 0.5\textwidth]{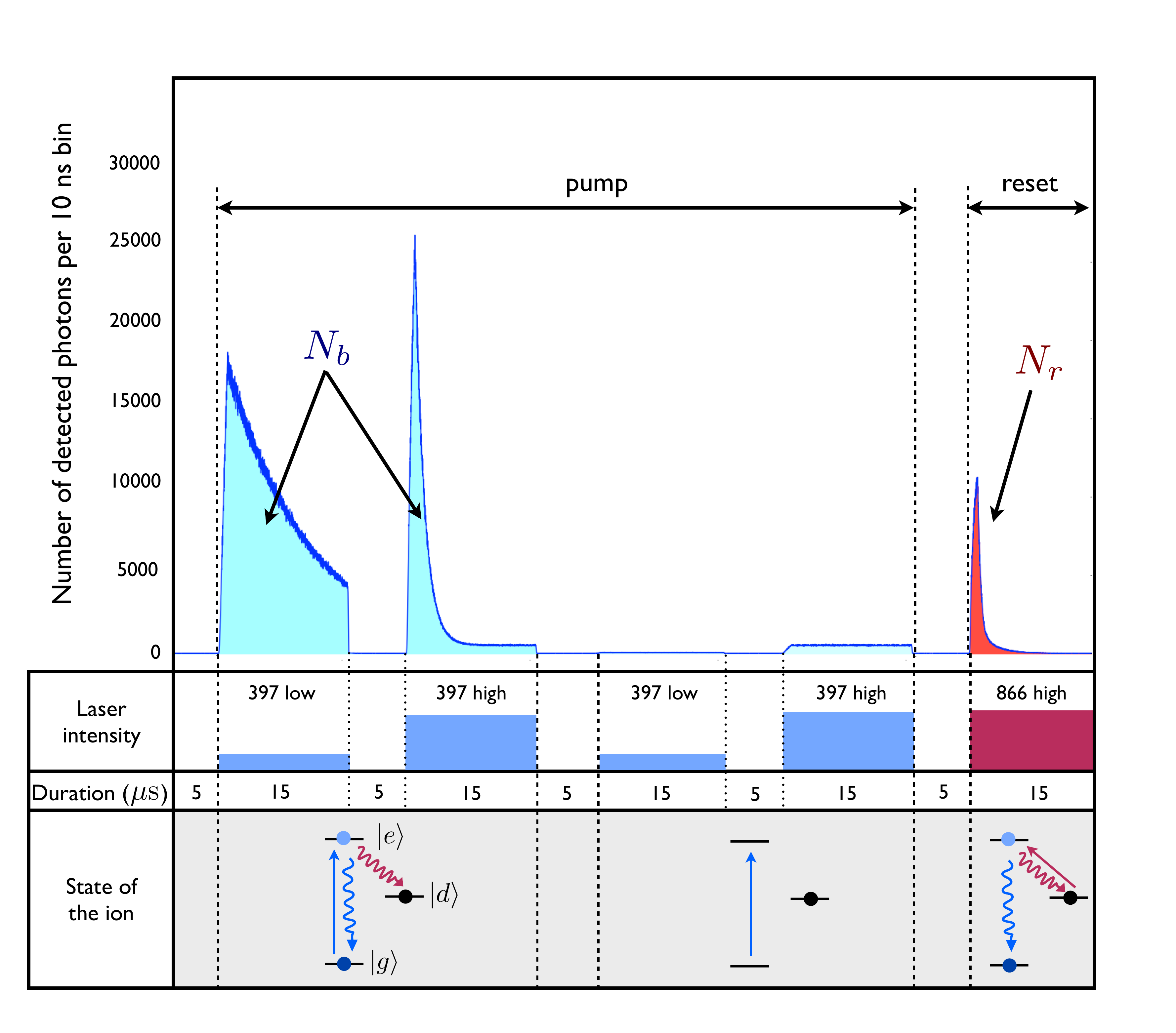}
\caption{\label{pulse}Fluorescence of blue photons at 397 nm collected from the ions after 50,000 detection cycles. The data is binned with a bin size of 10 ns. The first two 397 nm pump pulses transfer the ions to $\ket{d}$. Each pulse has a different intensity to minimize the count rate and the resultant error due to the PMT dead time. The next two 397 nm pump pulses determine the background counts. At the end of each detection cycle, we apply a reset pulse by illuminating the 866 nm light to pump all ions back to $\ket{g}$. $N_b$ is the total number of blue photons detected during the pump pulses. $N_r$ is the total number of blue photons detected during the reset pulse. The branching fraction can be determined from $p = N_b/(N_b+N_r)$.}
\end{center}
\end{figure}

Each measurement iteration consists of a Doppler cooling period of 5 ms followed by 50 detection cycles. Within each detection cycle, the 397 nm light is switched on, first, with low intensity and then with high intensity for 15~$\mu \text{s}$ each time. This ensures that the fluorescence count rate is kept low enough in order to minimize counting errors due to the PMT dead time. After the two 397 nm pulses, all the ions are pumped into $\ket{d}$. Then, the two pulses are repeated in order to measure the background scatterings of the laser. Finally, a 15~$\mu \text{s}$-long pulse of 866 nm returns the ions to $\ket{g}$. The detected photons are time-tagged and binned, as shown in Figure \ref{pulse}.

\section{Result and error analysis}

We repeat the measurement for a total of $\sim 10^8$ detection cycles. The total measurement time is $\sim$ 10 hours. We determine the number of photons emitted from the ions during the reset pulse, $N_r$, by subtracting the background recorded during the time where all lasers are off. The number of photons emitted from the ions during the 397 nm pump pulses, $N_b$, is found by taking the differences in the areas between the first two and the last two 397 nm laser pulses. The statistical error of the measurement is mostly due to photon number fluctuations of $N_r$. As a result, we measure the branching fractions for the 4$^{2}$P$_{1/2}$ to 4$^{2}$S$_{1/2}$ and 3$^{2}$D$_{3/2}$ to be \branching\  and \branchingsm, respectively.

The main systematic effects for the measurement are the residual birefringence in the detection optics, the dead time of the PMT, the finite lifetime of the 3$^2$D$_{3/2}$ state, and the light extinction ratio provided by the AOMs. The summary of the measurement uncertainties are presented in Table \ref{table:error}.  
\begin{table}[ht]
\centering
\begin{tabular}{c c c}
\hline\hline
\textbf{Effect} &    \textbf{Shift} & \textbf{Error} \\ [1.0ex] 
\hline
Photon counting statistical error					&-				&$5 \times 10^{-5}$		\\
Detection optics birefringence						&-				&$5 \times 10^{-5}$		\\
PMT dead time								&$ 7 \times 10^{-6}$ 		&$3 \times 10^{-6}$		\\
Lifetime of 3$^{2}$D$_{3/2}$ State 					&				&$2 \times 10^{-6}$		\\
Extinction ratio of AOMs 						&-				&$5 \times 10^{-6}$		\\
Off-resonant excitation to  4$^{2}$P$_{3/2}$ state			&$<1 \times 10^{-6}$		& -				\\
Finite measurement duration						&$15 \times 10^{-6}$		& $6 \times 10^{-6}$ 		\\
\hline
Total							&$22 \times 10^{-6}$		& $7 \times 10^{-5}$ 		\\
\hline
\hline
\end{tabular}
\caption{\label{table:error} List of measurement uncertainties.}
\end{table}

For the measurement to be insensitive to laser light polarizations, the detection optics should not distinguish between left and right circular polarizations. However, a small amount of residual birefringence in the detection path may lead to a varying detection efficiency. In order to experimentally check for this effect, we vary the direction of applied the magnetic field. We observe no variation in the branching fractions within the statistical uncertainty of the measurement.

The dead time of the PMT leads to undercounting of both $N_r$ and $N_b$. If the PMT is not saturated, however, it is possible to correct the count rate for this effect. The uncertainty of this correction is the result of not accurately knowing the dead time. Limiting the maximum count rate with multiple pulses of lower intensity allows one to significantly reduce the correction. 

Occasionally, the ion is excited off-resonantly to the 4$^2$P$_{3/2}$ level with the 397 nm laser, which may affect photon counting during the cycles when this excited state decays to the 3$^2$D$_{5/2}$ state. However, such events are sufficiently rare and do not affect the measurement results. The finite measurement duration accounts for the counts not recorded in the tails of the fluorescence peaks. Finally, the finite extinction ratio of the AOMs and RF switches accounts for a small residual coupling between the energy levels when the laser lights are switched off.

\section{Discussions and Summary}

\begin{figure}
\begin{center}
\includegraphics[width = 0.5\textwidth]{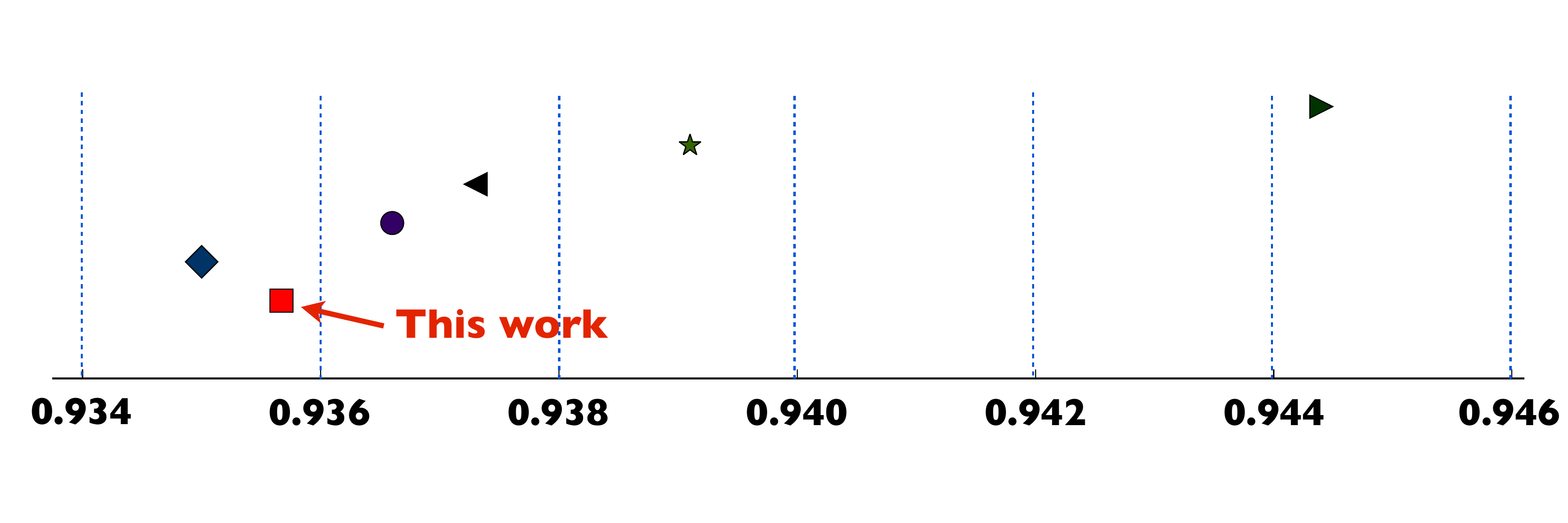}
\caption{\label{theory} Theoretical calculations of the branching fraction, $p$, of the 4$^{2}$P$_{1/2}$ state decay to the 4$^{2}$S$_{1/2}$ state for $^{40}$\Ca: ($\bullet$, $\blacktriangleright$) Liaw \textit{et al.} \cite{Liaw}, ($\star$) Guet \textit{et al.}  \cite{Guet}, ($\blacktriangleleft$) Sahoo \textit{et al.}  \cite{Sahoo}, ($\blacklozenge$) Arora \textit{et al.}  \cite{Arora}, compared to this work ($\blacksquare$).}
\end{center}
\end{figure}

Despite many theoretical calculations of the branching fractions of the excited state of $^{40}$\Ca \cite{Sahoo, Safronova, Arora, Guet, Liaw}, the only precise measurement of the branching fractions for the decay of the 4$^{2}$P$_{3/2}$ state has been reported in Ref.~\cite{Gerritsma}. However, the method is not applicable to the decay from the 4$^{2}$P$_{1/2}$ state due to its requirement of three decay channels for the excited state. The scheme used by Kurz \textit{et al.} \cite{Kurz} in \Ba can be applied to the 4$^{2}$P$_{1/2}$ state, but it requires an ultra-fast laser pulse which introduces additional systematic errors and complications in the experimental setup. Using the method presented in this work, we provide the first measurement of the branching fractions of the 4$^{2}$P$_{1/2}$ decay in \Ca at an uncertainty significantly smaller than the discrepancies between different theoretical calculations, as shown in Figure \ref{theory}.

Since the experimental scheme presented in our work does not require a narrow line-width laser addressing the quadrupole transition, it can be applied readily to perform precision measurement on many atomic species which have excited states with $J=1/2$ and a suitable long-lived state. In Au, both theoretical \cite{Migdalek} and experimental \cite{Hannaford, Fivet} studies on the atomic structure have been done and used to study the radiation spectrum from stars \cite{Barbuy}. In \Sc and \Sr, experimental measurements have yet to be done to verify theoretical calculations \cite{Nandy, Gallagher1967, Sansonetti}. Heavier ions like \Ba, \Yb and \Ra are especially interesting in precision spectroscopy due to their prospects of observing parity non-conservation effects \cite{Fortson, Versolato, Pal, Dzuba}, which can be investigated by measuring various branching fractions and lifetimes of different atomic levels. \Hg, where no study of the branching fractions have been done, also has a suitable level 
structure \cite{NIST}.

The statistical error in the measurement scheme presented here can be improved by either increasing the photon collection efficiency or the measurement time. Various experimental implementations have been demonstrated to increase photon collection efficiency to at least two orders of magnitude greater than that of our experimental setup \cite{Streed, VanDevender, Shu, Noek, Herskind}.

In summary, we present a method for measuring the branching fractions by simple photon counting and apply it to measure the branching fractions for the decay from the 4$^{2}$P$_{1/2}$ to the 4$^{2}$S$_{1/2}$ and 3$^{2}$D$_{3/2}$ states using an ion crystal. The scheme is shown to be insensitive to light intensity and frequency fluctuations. The $J = 1/2$ of the excited state provides additional insensitivity to magnetic field amplitude and orientation. The method can be applied readily to a variety of other atomic species with minimal experimental complication as it only relies on recording time resolved fluorescence and a minimal number of light sources.

We would like to acknowledge useful discussions with D. Budker. This work was supported by the NSF CAREER program grant \# PHY 0955650. MR was supported by an award from the Department of Energy (DOE) Office of Science Graduate Fellowship Program (DOE SCGF). 

\appendix

\end{document}